\documentclass[twocolumn,showpacs,preprintnumbers,amsmath,amssymb]{revtex4}

\usepackage{graphicx}
\usepackage{dcolumn}
\usepackage{bm}
\usepackage{graphicx} 
\newcommand{\etal}{{ \it et al. }}

\usepackage{amsmath,amsthm,amssymb,mathrsfs}
\usepackage{bm}
\usepackage{subfigure}

\newcommand{\D}{{\rm d}}

\begin{document}

\preprint{APS/123-QED}

\title{Dependence of critical current of spin transfer torque-driven 
        magnetization dynamics on free layer thickness}

\author{Tomohiro Taniguchi${}^{1,2}$}
\author{Hiroshi Imamura${}^{1}$\footnote{\vspace{-3ex}Corresponding author. Email address: h-imamura@aist.go.jp}}%
\affiliation{%
  ${}^{1}$
  Nanotechnology Research Institute, 
  National Institute of Advanced Industrial Science and Technology, 
  Central 2, 1-1-1, Umezono, Tsukuba, Ibaraki 305-8568, Japan \\
  ${}^{2}$
  Institute of Applied Physics, University of Tsukuba, Tsukuba, Ibaraki 305-8573, Japan
}%


\date{\today}

\begin{abstract}
 {
 The dependence of the critical current of spin transfer torque-driven
 magnetization dynamics on the free-layer thickness was studied by
 taking into account both the finite penetration depth of the transverse
 spin current and spin pumping.  We showed that the critical current
 remains finite in the zero-thickness limit of the free layer for both
 parallel and anti-parallel alignments.  We also showed that the
 remaining value of the critical current of parallel to anti-parallel
 switching is larger than that of anti-parallel to parallel switching.
 }
\end{abstract}

\pacs{Valid PACS appear here}

\maketitle


Spin transfer torque (STT)-driven magnetization dynamics is
a promising technique 
to operate spin-electronics devices 
such as a non-volatile magnetic random access memory (MRAM) 
and a microwave generator \cite{slonczewski96,berger96}. 
STT is the torque due to the transfer of the transverse 
(perpendicular to magnetization) spin angular momentum 
from the conducting electrons to the magnetization 
of the ferromagnetic metal. 
One of the most important quantities 
of STT-driven magnetization dynamics is 
the critical current over which 
the dynamics of the magnetization is induced. 
The typical value of the critical current density 
is on the order of $10^{6}-10^{8}$ [A/cm${}^{2}$] 
\cite{kiselev03,seki06,chen06}. 
Control of the value of the critical current 
is required to reduce the energy consumption of spin-electronics devices.


In Slonczewski's theory of STT \cite{slonczewski96}, 
the critical current of P-to-AP (AP-to-P) switching is 
expressed as \cite{sun00,grollier03}
\begin{equation}
  I_{\rm c}^{{\rm P}\to{\rm AP}({\rm AP}\to{\rm P})}
  =
  \frac{2eMSd}{\hbar\gamma\eta_{\rm P(AP)}}
  \alpha_{0}\omega_{\rm P(AP)}\ ,
  \label{eq:critical_current}
\end{equation}
where $e$ is the absolute value of the electron charge, 
$\hbar$ is the Dirac constant, 
and $M$, $\gamma$, $S$, $d$ and $\alpha_{0}$ are 
the magnetization, gyromagnetic ratio, cross section area, 
thickness and the intrinsic Gilbert damping constant 
of the free layer, respectively \cite{chen06}. 
$\omega_{\rm P(AP)}$ is 
the angular frequency of the magnetization around 
the equilibrium point. 
The coefficient $\eta_{\rm P,AP}$ characterizes the strength of STT, 
and depends only on the relative angle of the magnetizations of 
the fixed and free layer 
\cite{slonczewski96,sun00,grollier03}. 
According to Eq. (\ref{eq:critical_current}), 
the critical current vanishes 
in the zero-thickness limit of the free layer, $d\!\to\!0$.


\begin{figure}
  \centerline{\includegraphics[width=0.95\columnwidth]{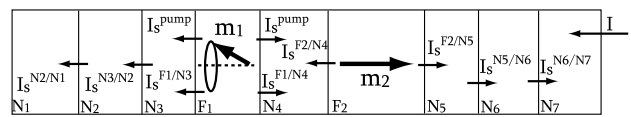}}
  \caption{
    The schematic view of the nonmagnetic(N) / ferromagnetic(F) multilayer. 
    $I$ and $\mathbf{I}_{s}^{\rm pump}$ are 
    the electric current and pumped spin current, respectively. 
    $\mathbf{I}_{s}^{{\rm N}_{i}({\rm F}_{k})/{\rm N}_{j}}$ is the spin current 
    induced by the spin accumulations in each layer. 
    $\mathbf{m}_{k}$$(k \! = \! 1,2)$ is the unit vector pointing the
    direction of the magnetization of the F${}_{k}$ layer. 
  }
  \vspace{-2ex}
  \label{fig:fig1}
\end{figure}


However, recently, Chen \etal \cite{chen06} reported that 
the critical current of STT-driven magnetization dynamics of a CPP-GMR
spin valve remains finite even in the zero-thickness limit of the free
layer. 
What are missed in the above naive considerations based on Slonczewski's
theory are the effects of the finite penetration depth of the transverse spin
current, $\lambda_{\rm t}$,  \cite{zhang02,zhang04,taniguchi08a} and of 
spin pumping \cite{mizukami02b,tserkovnyak02,tserkovnyak03,taniguchi07}.
We investigated the critical current of 
STT-driven magnetization switching from 
AP to P alignment by taking into account 
both the finite penetration depth of the transverse spin current 
and the spin pumping, and showed that 
the critical current remains finite 
in the zero-thickness limit of the free layer \cite{taniguchi08b}. 
We also showed that 
the remaining value of the critical current is 
mainly determined by spin pumping. 
Although our results \cite{taniguchi08b} agree well with the
experimental results of Chen \etal \cite{chen06}, we investigated only
the critical current of AP-to-P switching, $I_{\rm c}^{{\rm AP}\to{\rm P}}$. 
For the manipulation of spin-electronics devices, 
the thickness dependence of the critical current 
of P-to-AP switching, $I_{\rm c}^{{\rm P}\to{\rm AP}}$, 
should also be investigated.


In this paper, we study the critical current of STT-driven magnetization
switching  both from P to AP alignment and from AP to P alignment 
by taking into account both 
the finite penetration depth of the transverse spin current and the spin pumping. 
We show that both critical currents, 
$I_{\rm c}^{{\rm P}\to{\rm AP}}$ and $I_{\rm c}^{{\rm AP}\to{\rm P}}$, 
remain finite in the zero-thickness of the free layer. 
We also show that 
$I_{\rm c}^{{\rm P}\to{\rm AP}}$ is larger than $I_{\rm c}^{{\rm AP}\to{\rm P}}$ 
over the whole range of the free layer thickness, 
and thus, the remaining value of $I_{\rm c}^{{\rm P}\to{\rm AP}}$ is 
larger than that of $I_{\rm c}^{{\rm AP}\to{\rm P}}$. 
The difference between the remaining values of the critical currents,
$I_{\rm c}^{{\rm P}\to{\rm AP}}$ and $I_{\rm c}^{{\rm AP}\to{\rm P}}$,
can be explained by considering how the strength of STT, $\eta$, depends
on the magnetic alignment. 


A schematic view of the system we consider is shown in Fig. \ref{fig:fig1}. 
Two ferromagnetic layers (F${}_{1}$ and F${}_{2}$) are sandwiched 
by the nonmagnetic layers N${}_{i}$ $(i=1-7)$. 
The F${}_{1}$ and F${}_{2}$ layers correspond to 
the free and fixed layers, respectively. 
$\mathbf{m}_{k}$ $(k=1,2)$ is the unit vector pointing in the direction 
of the magnetization of the F${}_{k}$ layer. 
$I$ is the electric current flowing perpendicular to the film plane. 


The electric current and pumped spin current 
at the F${}_{k}$/N${}_{i}$ interface (into N${}_{i}$) 
is obtained by using the circuit theory \cite{tserkovnyak02,brataas01}:
\begin{align}
 &
 I^{{\rm F}_{k}/{\rm N}_{i}}
 \!=\!
 \frac{eg}{2h}
 \left[
 2(\mu_{{\rm F}_{k}}-\mu_{{\rm N}_{i}})
 \!+\!
 p\mathbf{m}_{k}\!\cdot\!(\bm{\mu}_{{\rm F}_{k}}-\bm{\mu}_{{\rm N}_{i}})
 \right]\ ,
 \label{eq:electric_current}
 \\
 &
 \mathbf{I}_{s}^{\rm pump}
 \!=\!
 \frac{\hbar}{4\pi}
 \left(
    g_{\rm r}^{\uparrow\downarrow}
 \mathbf{m}_{1}\!\times\!
 \frac{\D\mathbf{m}_{1}}{\D t}
 \!+\!
 g_{\rm i}^{\uparrow\downarrow}
 \frac{\D\mathbf{m}_{1}}{\D t}
 \right)\ ,
 \label{eq:pump_current}
\end{align}
where $h\!=\!2\pi\hbar$ is the Planck constant, 
$g\!=\!g^{\uparrow\uparrow}\!+\!g^{\downarrow\downarrow}$ is the sum of 
the spin-up and spin-down conductances, 
$p\!=\!(g^{\uparrow\uparrow}\!-\!g^{\downarrow\downarrow})/(g^{\uparrow\uparrow}\!+\!g^{\downarrow\downarrow})$ 
is the spin polarization of the conductances, 
and $g_{\rm r(i)}$ is the real (imaginary) part of the mixing conductance. 
$\mu_{{\rm N}_{i},{\rm F}_{k}}$ and $\bm{\mu}_{{\rm N}_{i},{\rm F}_{k}}$ are 
the charge and spin accumulation, respectively. 
The spin current at each F${}_{k}$/N${}_{i}$ and N${}_{i}$/N${}_{j}$ interface 
(into N${}_{i}$) is given by \cite{taniguchi08a,brataas01} 
\begin{align}
 \!\!\!\!
 &\mathbf{I}_{s}^{{\rm F}_{k}/{\rm N}_{i}}
 \!=
  \frac{1}{4\pi}\!
  \left[
    g
    \left\{\!
     p(\mu_{{\rm F}_{k}}\!-\!\mu_{{\rm N}_{i}})
     \!+\!
     \frac{1}{2}
     \mathbf{m}_{k}\!\cdot\!(\bm{\mu}_{{\rm F}_{k}}\!-\!\bm{\mu}_{{\rm N}_{i}})\!
    \right\}
    \mathbf{m}_{k}
  \right.\nonumber \\
  &\hspace{3em}-
    g_{\rm r}^{\uparrow\downarrow}
    \mathbf{m}_{k}\!\times\!(\bm{\mu}_{{\rm N}_{i}}\!\times\!\mathbf{m}_{k})
    \!-\!
    g_{\rm i}^{\uparrow\downarrow}
    \bm{\mu}_{{\rm N}_{i}}\!\times\!\mathbf{m}_{k}
 \nonumber\\
  &\hspace{3em} + \left.
    t_{\rm r}^{\uparrow\downarrow}
    \mathbf{m}_{k}\!\times\!(\bm{\mu}_{{\rm F}_{k}}\!\times\!\mathbf{m}_{k})
    \!+\!
    t_{\rm i}^{\uparrow\downarrow}
    \bm{\mu}_{{\rm F}_{k}}\!\times\!\mathbf{m}_{k}
  \right]\ ,
  \label{eq:spin_current_FN}\\
 &\mathbf{I}_{s}^{{\rm N}_{i}/{\rm N}_{j}}
  \!=\!
  -\frac{g_{{\rm N}_{i}/{\rm N}_{j}}}{4\pi}
  (\bm{\mu}_{{\rm N}_{i}}\!-\!\bm{\mu}_{{\rm N}_{j}})\ ,
  \label{eq:spin_current_NN}
\end{align}
where $t_{\rm r(i)}^{\uparrow\downarrow}$ is the real (imaginary) part 
of the transmission mixing conductance at the F${}_{k}$/N${}_{i}$ interface 
and $g_{{\rm N}_{i}/{\rm N}_{j}}$ is the conductance of the one spin channel 
at the N${}_{i}$/N${}_{j}$ interface. 


The spin accumulations in the N and F layer obey 
the diffusion equation \cite{zhang02,taniguchi08a,valet93}. 
The spin accumulation in the N layer, $\bm{\mu}_{\rm N}$, decays exponentially 
with the spin diffusion length $\lambda_{\rm sd(N)}$. 
The longitudinal and transverse spin accumulations in the F layer are
defined as $(\mathbf{m}\cdot\bm{\mu}_{\rm F})\mathbf{m}$ and 
$\mathbf{m}\times(\bm{\mu}_{\rm F}\times\mathbf{m})$, respectively.
The longitudinal and transverse spin accumulations decay exponentially
with the spin diffusion length $\lambda_{\rm sd(F_{L})}$ 
and with the penetration depth of the transverse spin current
$\lambda_{\rm t}$, respectively. 


The total spin currents across the N${}_{3}$/F${}_{1}$ and F${}_{1}$/N${}_{4}$ interfaces, 
i.e., $\mathbf{I}_{s}^{(1)}\!=\!\mathbf{I}_{s}^{\rm pump}\!+\!\mathbf{I}_{s}^{\rm F_{1}/N_{3}}$ 
and $\mathbf{I}_{s}^{(2)}\!=\!\mathbf{I}_{s}^{\rm pump}\!+\!\mathbf{I}_{s}^{\rm F_{1}/N_{4}}$, exert the torque 
$\bm{\tau}\!=\mathbf{m}_{1}\!\times\![(\mathbf{I}_{s}^{(1)}\!+\!\mathbf{I}_{s}^{(2)})\!\times\!\mathbf{m}_{1}]$ 
on the magnetization $\mathbf{m}_{1}$. 
In order to obtain the spin current $\mathbf{I}_{s}^{(1,2)}$, 
we solve the diffusion equation of spin accumulation in each layer. 
The boundary conditions are as follows. 
We assume that the thicknesses of the N${}_{1}$ and N${}_{7}$ layer 
are much larger than their spin diffusion length, 
and that the spin current is zero at the outer boundary 
of the N${}_{1}$ and N${}_{7}$ layer. 
We also assume that the spin current is continuous at all interfaces 
and that the electric current is constant 
through the entire structure. 


The torque $\bm{\tau}$ modifies the Landau-Lifshitz-Gilbert (LLG) equation 
of magnetization $\mathbf{m}_{1}$ as \cite{tserkovnyak02} 
\begin{equation}
\begin{split}
  \frac{\D\mathbf{m}_{1}}{\D t}
  \!=\! &
  -\!\gamma
  \mathbf{m}_{1}\!\times\!\mathbf{B}_{\rm eff}
  \!+\!
  \frac{\gamma}{MSd}
  \bm{\tau}
  \!+\!
  \alpha_{0}
  \mathbf{m}_{1}
  \!\times\!
  \frac{\D\mathbf{m}_{1}}{\D t}
\\
  &=\!
  -\!\gamma_{\rm eff}
  \mathbf{m}_{1}\!\times\!\mathbf{B}_{\rm eff}
  \!+\!
  \frac{\gamma_{\rm eff}}{\gamma}
  (\alpha_{0}+\alpha^{'})
  \mathbf{m}_{1}\!\times\!\frac{\D\mathbf{m}_{1}}{\D t}\ ,
\end{split}
\end{equation}
where $\mathbf{B}_{\rm eff}$ is the effective magnetic field, and 
$\alpha^{'}\!=\!\alpha_{c}\!+\!\alpha_{\rm pump}$ is 
the enhancement of the Gilbert damping constant. 
The enhancement $\alpha_{c}$ is proportional 
to the electric current 
and independent of the pumped spin current. 
The enhancement $\alpha_{\rm pump}$ represents 
the contribution from the pumped spin current 
and is independent of the electric current. 
The enhancement of the gyromagnetic ratio, 
$\gamma_{\rm eff}/\gamma$, is a function 
of both the electric current and the pumped spin current.


The critical current of the STT-driven magnetization dynamics 
is defined by the electric current 
that satisfies the condition, 
$\alpha_{0}\!+\!\alpha_{c}\!+\!\alpha_{\rm pump}\!=\!0$, and given by 
\begin{equation}
  I_{\rm c}^{{\rm P}\to{\rm AP}({\rm AP}\to{\rm P})}
  \!=\!
  \frac{2eMSd}{\hbar\gamma\tilde{\eta}_{\rm P(AP)}}
  (\alpha_{0}\!+\!\alpha_{\rm pump})
  \omega_{\rm P(AP)}\ ,
  \label{eq:critical_current_TT}
\end{equation}
where the coefficient $\tilde{\eta}_{\rm P,AP}$ 
characterizes the strength of STT due to the electric current, 
and is determined by the diffusion equations of the spin accumulations. 
Thus, $\tilde{\eta}_{\rm P,AP}$ is the function of 
$d/\lambda_{\rm sd(F_{L})}$, $d/\lambda_{\rm t}$ and 
the relative angle of the magnetizations of the F${}_{1}$ and F${}_{2}$ layers.


\begin{figure}
  \centerline{\includegraphics[width=0.8\columnwidth]{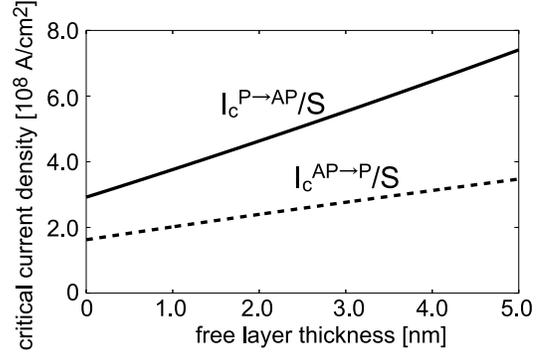}}
  \caption{
  The critical current densities of 
  P-to-AP switching $(I_{\rm c}^{{\rm P}\to{\rm AP}}/S)$ and 
  AP-to-P switching $(I_{\rm c}^{{\rm AP}\to{\rm P}}/S)$ in 
  STT-driven magnetization dynamics 
  are shown against the free layer thickness. 
  }
  \vspace{-2ex}
  \label{fig:fig2}
\end{figure}


We performed numerical calculation to obtain the critical currents
$I_{\rm c}^{{\rm P}\to{\rm AP}}$ and $I_{\rm c}^{{\rm AP}\to{\rm P}}$.
The system consists of nine layers as shown in Fig. \ref{fig:fig1}, 
where F${}_{1}$ and F${}_{2}$ are Co, 
N${}_{1}$, N${}_{3}$, N${}_{4}$, N${}_{5}$ and N${}_{7}$ are Cu, 
and N${}_{2}$ and N${}_{6}$ are Pt. 
The thicknesses of the N${}_{3}$, N${}_{4}$ and N${}_{5}$ layers are 10 nm, 
the thicknesses of the N${}_{2}$ and N${}_{6}$ layers are 3 nm 
and the thickness of the F${}_{2}$ layer is 12 nm \cite{chen06}. 
The thickness of the N${}_{1}$ and N${}_{7}$ layers are taken to be 10 $\mu$m. 
The spin diffusion length of Cu and Pt are 
1000 and 14 nm, respectively \cite{bass07}. 
The conductance at the Cu/Pt interface 
is 35 nm${}^{-2}$ \cite{bass07}. 
The magnetization, 
the intrinsic Gilbert damping constant and
the gyromagnetic ratio of Co are 
0.14 T, 0.008 and $1.89 \! \times \! 10^{11}$ Hz/T, 
respectively \cite{chen06,beaujour06}. 
The polarization $p$ is taken to be $0.46$ for Co \cite{bass07}. 
The spin diffusion length of Co is 40 nm \cite{bass07}. 
The penetration depth of the transverse spin current 
of Co is 4.2 nm \cite{zhang04,taniguchi08b}. 
The conductances at the Co/Cu interface, 
$g/S$, $g_{\rm r}^{\uparrow\downarrow}/S$ and $g_{\rm i}^{\uparrow\downarrow}/S$, are 
50, 27 and 0.4 nm${}^{-2}$, respectively \cite{tserkovnyak02,tserkovnyak03,brataas01}. 
We assume that $t_{\rm r} \! = \! t_{\rm i}$ 
where $t_{\rm r,i}/S$ at the Co/Cu interface is taken to be 6.0 nm${}^{-2}$. 
The angular frequency is 
$\omega_{\rm P(AP)} \! = \! \gamma[B_{\rm appl} \!-\!(+)4\pi M]$ 
where the strength of the applied magnetic field 
$B_{\rm appl}$ is 7 T \cite{chen06}.


\begin{figure}
  \centerline{\includegraphics[width=0.8\columnwidth]{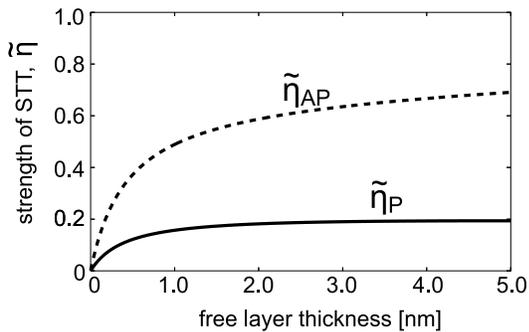}}
  \caption{
      The coefficient $\tilde{\eta}$ 
      in the P state $(\tilde{\eta}_{\rm P})$ and AP state $(\tilde{\eta}_{\rm AP})$, 
      against the free layer thickness.
  }
  \vspace{-2ex}
  \label{fig:fig3}
\end{figure}


Figure \ref{fig:fig2} shows 
the dependence of the critical current density of Eq. 
\eqref{eq:critical_current_TT} for 
P-to-AP switching, $I_{\rm c}^{{\rm P}\to{\rm AP}}/S$, and 
AP-to-P switching, $I_{\rm c}^{{\rm AP}\to{\rm P}}/S$, 
on the free layer thickness, $d$. 
As shown in Fig. \ref{fig:fig2}, 
both $I_{\rm c}^{{\rm P}\to{\rm AP}}$ and $I_{\rm c}^{{\rm AP}\to{\rm P}}$ 
remain finite in the zero-thickness limit of the free layer. 
We show that the critical current $I_{\rm c}^{{\rm P}\to{\rm AP}}$ 
is larger than $I_{\rm c}^{{\rm AP}\to{\rm P}}$ 
over the whole range of the free layer thickness, 
and thus, the remaining value of $I_{\rm c}^{{\rm P}\to{\rm AP}}$ 
is larger than that of $I_{\rm c}^{{\rm AP}\to{\rm P}}$. 
As shown in Ref.  \cite{taniguchi08b},
the remaining value of the critical current 
is mainly determined by spin pumping.
It should be noted that the magnitude of the enhancement of the Gilbert
damping constant due to spin pumping, $\alpha_{\rm pump}$, 
is the same for both P-to-AP switching and AP-to-P switching
\cite{tserkovnyak03,taniguchi07}. 
Thus, the fact that the remaining values
$I_{\rm c}^{{\rm P}\to{\rm AP}}$ and $I_{\rm c}^{{\rm AP}\to{\rm P}}$ 
are different from each other implies that the strength of STT,
$\tilde{\eta}_{\rm P,AP}$, depends on the alignment of the magnetizations. 
As shown in Fig. \ref{fig:fig3}, $\tilde{\eta}_{\rm P,AP}$ decreases
with a decreasing free layer thickness.
On the other hand, the number of localized magnetic moments in the
free layer, and therefore the STT per magnetic moment, is inversely
proportional to the free layer thickness $d$.
According to Eq. (\ref{eq:critical_current_TT}), 
the remaining value of the critical current is proportional to 
$(\tilde{\eta}_{\rm P,AP}/d)^{-1}$ with $d\!\to\!0$, 
where $\tilde{\eta}_{\rm P}/d\simeq 0.44$ nm${}^{-1}$ and 
$\tilde{\eta}_{\rm AP}/d\simeq 1.47$ nm${}^{-1}$ 
in the limit of $d\!\to\!0$ are estimated 
by Fig. \ref{fig:fig3}. 
Thus, the remaining value of $I_{\rm c}^{{\rm P}\to{\rm AP}}$ is larger 
than that of $I_{\rm c}^{{\rm AP}\to{\rm P}}$.


In summary, we studied the critical current of STT-driven magnetization dynamics 
by taking into account the finite penetration depth of the transverse spin current 
and spin pumping for both P and AP magnetic alignments.
We showed that the critical current remains finite in the zero thickness
limit of the free layer for both P-to-AP  and AP-to-P switching.
We also showed that the critical current for P-to-AP switching is larger 
than that for AP-to-P switching over the whole range of the free layer thickness. 


The authors would like to acknowledge the valuable discussions 
they had with K. Matsushita, J. Sato and N. Yokoshi. 
This work was supported by JSPS and NEDO.



\end{document}